\documentclass[]{aa}  
\usepackage{graphicx}
\usepackage{epsfig}
\usepackage{txfonts}
\begin{document}
\title{Complicated variations of early optical afterglow of GRB\,090726}
\author{V.~\v{S}imon \inst{1}, C.~Pol\'{a}\v{s}ek \inst{1}, M.~Jel\'{\i}nek \inst{2}, 
R.~Hudec \inst{1}, J.~\v{S}trobl \inst{1} }  
\offprints{V.~\v{S}imon: simon@asu.cas.cz} 
\institute{Astronomical Institute, Academy of Sciences of the Czech 
Republic, 251~65~Ond\v{r}ejov, Czech Republic \and Instituto de Astrof\'{\i}sica 
de Andaluc\'{\i}a CSIC, Apartado de Correos, 3.004, E-18.080 Granada, Spain }
\date{Received date; accepted date} 
\authorrunning{\v{S}imon et al.} 
\titlerunning{Early optical afterglow of GRB\,090726}

\abstract
{}
{We report on a detection of an early rising phase of optical afterglow (OA) 
of a long  GRB\,090726. We resolve a complicated profile of the optical light 
curve. We also investigate the relation of the  optical and X-ray emission of 
this event. }
{We make  use of  the optical photometry  of this  OA  obtained  by the 0.5~m
telescope of AI~AS~CR, supplemented by the data obtained by  other observers,
and the X-ray {\it Swift}/XRT data.   }
{The  optical  emission  peaked  at  $\sim17.5$~mag($R$)  at $t-T_{0} \approx
~500$~s. We  find  a complex  profile  of  the  light  curve during the early
phase of  this  OA: an  approximately power-law rise, a rapid transition to a
plateau, a weak  flare superimposed  on the  center of  this  plateau,  and a
slowly  steepening early decline followed  by a power-law  decay. We  discuss 
several  possibilities  to explain  the short  flare on  the  flat top of the 
optical light  curve at $t-T_{0}\approx500$~s; activity of the central engine 
is favored although reverse shock cannot be ruled out. We show that power-law 
outflow with $\Theta_{\rm obs}/\Theta_{\rm c}>2.5$ is the best case for OA of 
GRB\,090726. The initial Lorentz factor is $\Gamma_{0}\approx230-530$ in case 
of propagation  of the  blast wave in a homogeneous medium, while propagation 
of this wave in  a wind  environment  gives  $\Gamma_{0} \approx 80-300$. The 
value  of $\Gamma_{0}$ in  GRB\,090726  thus falls into the lower half of the 
range observed in GRBs and it may even lie  on the  lower  end. We  also show 
that both the optical and X-ray  emission decayed simultaneously and that the 
spectral  profile  from  X-ray to the optical band did not vary. This is true 
for both the time period before and after the break in the X-ray light curve. 
This  break  can be  regarded  as achromatic. The  available  data  show that 
neither  the  dust  nor  the  gaseous  component  of  the  circumburst medium 
underwent any evolution during the decay of this OA, that is after $t-T_{0} < 
3000$~s. We  also show that this OA belongs to the least luminous ones in the 
phase of its power-law decay  corresponding to that observed for the ensemble 
of OAs of long GRBs.  }
{}


\maketitle

\section{Introduction    \label{int}   }

    GRB\,090726 was a long gamma-ray burst (GRB) localized by {\it Swift}/BAT 
on the  26$^\mathrm{th}$~July  2009  at 22:42:27.8\,UT. It displayed a single 
peak  profile,  with  $T_{90}$  (15-350~keV)  $= 67 \pm 15$~s. Fluence in the 
15--150~keV  band  was  (8.6$\pm$1.0)$\times 10^{-7}$~erg~cm$^{-2}$~s$^{-1}$. 
The  1-s  peak  flux  was  0.7$\pm$0.2 photon cm$^{-2}$ s$^{-1}$ (Page et al. 
2009). 

     The GRB  coordinates  were  available  via  GCN  within  65\,s, although 
because of  an  Earth-limb  constraint  {\it Swift} could not follow-up until 
$\sim$3.6\,ks  after  the  trigger. Redshift  of  this  GRB is $z = 2.71$, as 
determined  from the lines  in optical  spectrum of the OA (Fatkhullin et al. 
2009).

     In this letter, we  report  the early  observations  of optical afterlow 
(OA) of this  GRB  performed  by  the  robotic  0.5\,m telescope (D50) of the 
Astronomical Institute  of  AS~CR  in Ond\v{r}ejov, Czech Republic. The first 
image of the  D50  started at  22:45:41.0\,UT, i.e. 194~s  after the trigger. 
The telescope  was taking images until 23:35:35\,UT (for 50 minutes), until a 
pointing limit was reached. On all of the 107 unfiltered, 20\,s exposures the 
OA can  be  detected, although, to  improve  the  signal  to  noise ratio, we 
co-added some of the images.

\section{Observations and data analysis}  \label{source}

     D50 is a robotic, 500/1975\,mm  Newtonian telescope, located at AI~AS~CR 
in Ond\v{r}ejov. It  is equipped with a CCD camera FLI~IMG 4710 (CCD chip E2V 
47-10, mid-band coating, 1024$\times$1024  pixels),  focuser  FLI DF-2, and a 
field corrector TeleVue Paracorr PSB-11000. It is controlled by RTS2 software 
(Kub\'{a}nek  et  al. 2006) and  makes use of astrometric code by Lang et al. 
(2009). The  field  of view  is $20\times20$~arcmin. In order to reach deeper 
magnitudes, unfiltered  observations were carried out. Their peak sensitivity 
is close to the $R$ filter. Exposure time of 20~s was used for each CCD image.

\begin{table}
\caption[ ]{Calibration stars used, together  with their USNO-A2.0 
derived  magnitudes. Denoted  error  estimates ($1\sigma$) are for 
relative brightnesses within this frame, external error~-- binding 
to USNO system ($\sim$0.04\,mag) and error of the USNO calibration 
itself ($\sim$0.2\,mag)~-- are much larger than this.  }
\label{calib}
\begin{center}
\[
\begin{array}{lcclc}
\hline
\noalign{\smallskip}
$id$ &       $mag$    & \qquad & $id$ &        $mag$    \\
\hline
1    & 14.203\pm0.001 &        & 5    &  15.854\pm0.006 \\
2    & 16.602\pm0.010 &        & 6    &  15.980\pm0.006 \\
3    & 16.689\pm0.011 &        & 7    &  17.139\pm0.015 \\
4    & 16.012\pm0.007 &        & 8    &  17.185\pm0.018 \\
\noalign{\smallskip}
\hline
\end{array}
\]
\end{center}
\end{table}

                          Aperture photometry using {\tt phot} routine within 
IRAF/DAOPhot\footnote{IRAF is  distributed  by the National Optical Astronomy 
Observatory, which  is  operated  by  the  Association  of  Universities  for 
Research in Astronomy, Inc., under co-operative agreement with  the  National 
Science Foundation.} was used on the OA and 8 comparison stars. The reference 
grid of  these  8 stars  was then calibrated with brightnesses from USNO-A2.0 
(Monet  et  al.  1998)  using robust fitting  algorithms  to  avoid  possible 
influence of star variability. The values we obtained for brightnesses of the 
calibration stars are  in  Table~\ref{calib}. As we  have no afterglow  color 
information, we  cannot transform the  CR  (clear, $R$-calibrated) magnitudes 
into real $R$ band. Because  the airmass changed only between 1.180 and 1.124 
during the  observation, the atmospheric  color effects should be negligible. 
CR magnitude  of the  OA differs  from  the $R$ band  only  by the difference 
between  the OA  color  and  the mean  color  of the calibration stars. If we 
assume  the  (typically) constant color of the OA, also this difference stays 
constant. The data are given in Table~\ref{obs}. 

\begin{figure}
\centering
 \includegraphics[width=0.48\textwidth]{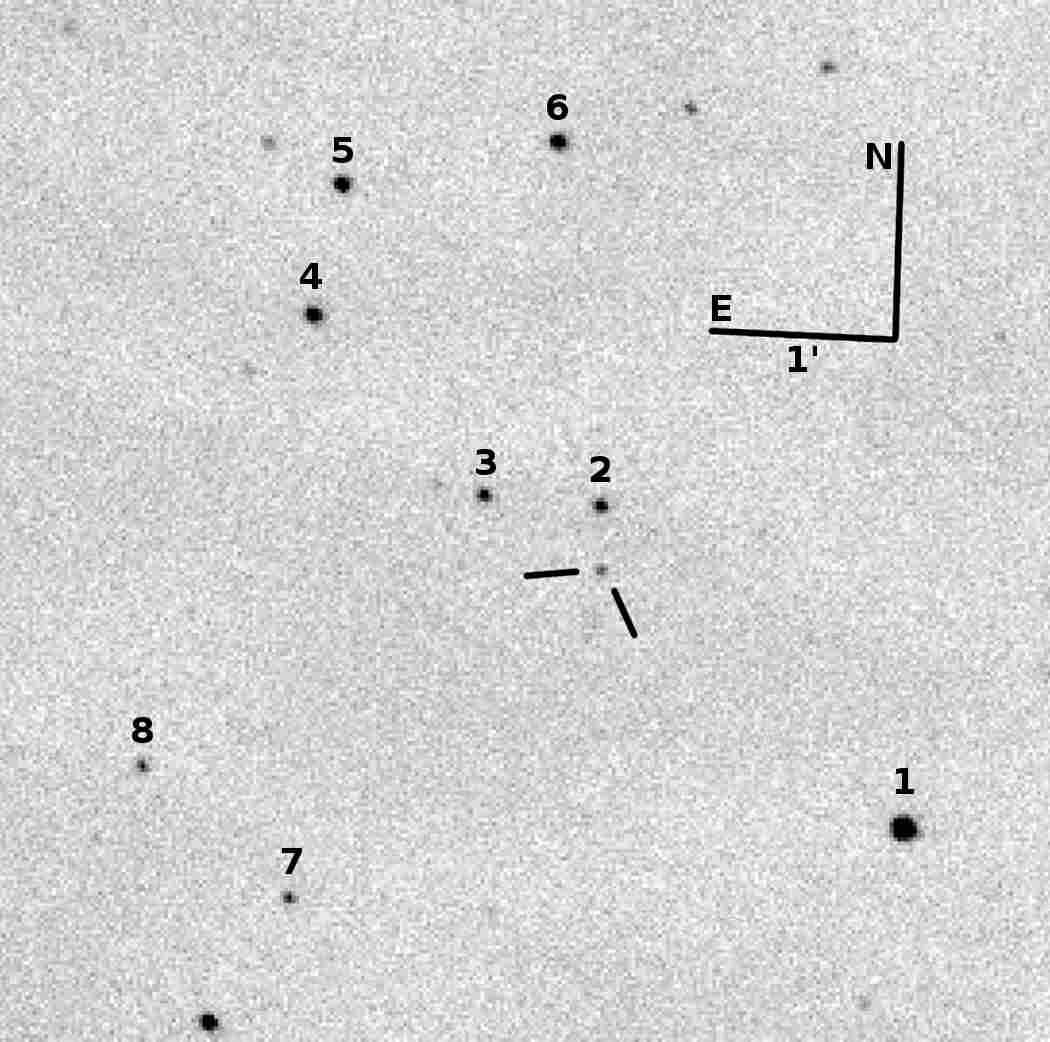} 
\caption{CCD image of OA of GRB\,090726 secured with the 
D50~cm  telescope. It  was obtained close to the time of 
the peak brightness of this OA, at $t-T_{0}\approx670$~s. 
Only  a part  of the original image is shown. OA and the 
comparison stars are marked. See Sect.\,\ref{source} for 
details.   }
\label{CCD}
\end{figure}

\begin{figure}[t!,h]
\centering
 \includegraphics[width=0.45\textwidth]{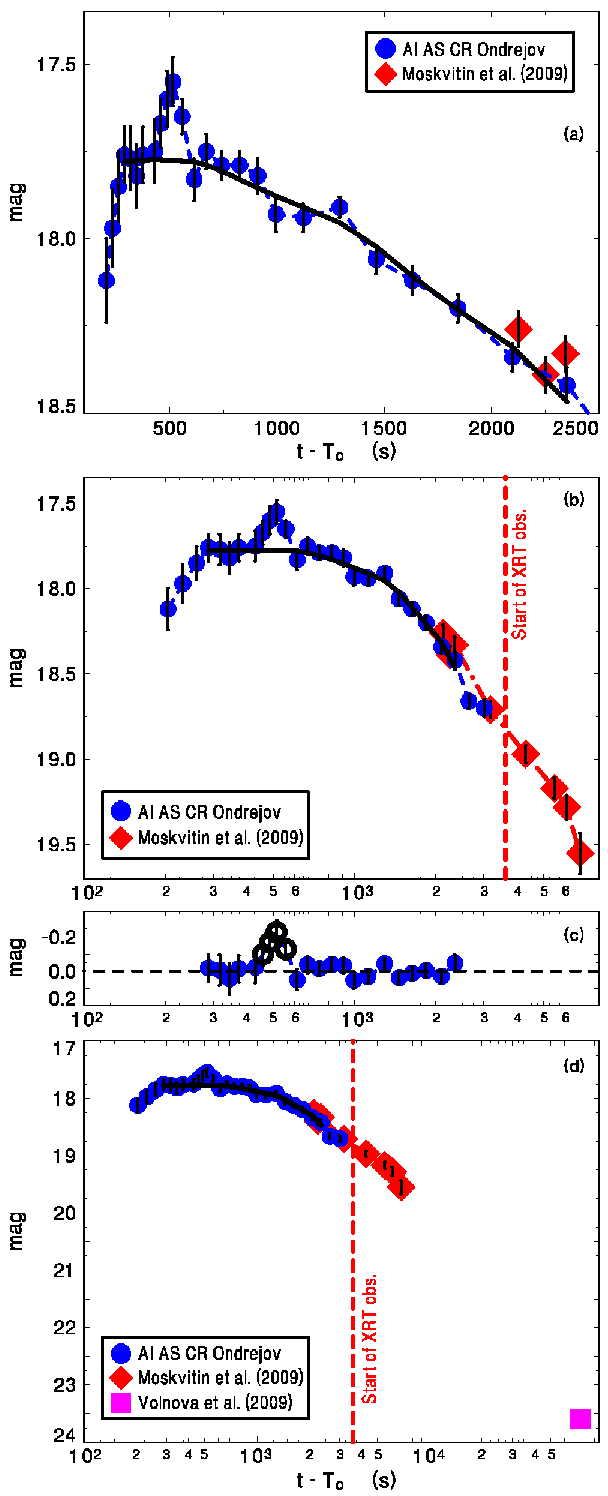} 
\caption{Light  curve of  OA of GRB\,090726 in  the red 
spectral  region. The  data coming from  the individual 
observers are resolved. HEC13 fit to  Ond\v{r}ejov data 
is marked  by  the  smooth, solid  line. {\bf  (a)}~The 
very early  phase  plotted  with a  linear scale on the 
abscissa. {\bf (b)} The early phase with  a logarithmic 
scale  of the  abscissa. {\bf (c)} Residuals  of  HEC13 
fit. The  open  circles  are  highlighting  the  flare. 
{\bf  (d)}  The  first  day  of  OA.  A  vertical  line 
denotes the start  of {\it Swift}/XRT observations. The 
error bars represent the uncertainties at the $1\sigma$ 
level. See Sect.\,\ref{source} for details.   }
\label{opt}
\end{figure}

\begin{table}
\caption[ ]{Observing log of Ond\v{r}ejov data. Time 
elapsed  from  GRB trigger  in seconds  is marked as 
$t-T_{0}$  while $E$  refers to the exposure time in 
seconds  (some images  are  co-added). Magnitude  is 
denoted as mag, including its $1\sigma$ error.  }
\begin{center} 
\label{obs}
\[
\begin{array}{llcccllc}
\hline
\noalign{\smallskip}
t-T_{0} &  E   &      $mag$      &  &  & t-T_{0} &  E   &      $mag$      \\
\hline
204.2  & 20    &  18.12\pm0.12   &  &  & 741.3  &  77   &  17.79\pm0.04   \\
232.2  & 20    &  17.97\pm0.11   &  &  & 825.5  &  76   &  17.79\pm0.04   \\
260.7  & 20    &  17.85\pm0.10   &  &  & 910.5  &  76   &  17.82\pm0.05   \\
288.7  & 20    &  17.76\pm0.08   &  &  & 995.5  &  77   &  17.93\pm0.05   \\
317.2  & 20    &  17.77\pm0.09   &  &  & 1123.0 &  162  &  17.94\pm0.04   \\
345.2  & 20    &  17.82\pm0.09   &  &  & 1292.2 &  161  &  17.91\pm0.03   \\
373.8  & 20    &  17.76\pm0.08   &  &  & 1462.1 &  162  &  18.06\pm0.04   \\
430.3  & 20    &  17.75\pm0.09   &  &  & 1631.1 &  161  &  18.12\pm0.04   \\
458.4  & 20    &  17.67\pm0.07   &  &  & 1843.1 &  246  &  18.20\pm0.04   \\
486.8  & 20    &  17.60\pm0.08   &  &  & 2097.6 &  247  &  18.34\pm0.04   \\
514.8  & 20    &  17.55\pm0.07   &  &  & 2351.7 &  246  &  18.42\pm0.05   \\
557.3  & 48    &  17.65\pm0.05   &  &  & 2648.2 &  331  &  18.66\pm0.04   \\
613.8  & 48    &  17.83\pm0.06   &  &  & 3015.1 &  387  &  18.70\pm0.05   \\
670.3  & 48    &  17.75\pm0.05   &  &  &        &       &                 \\
\noalign{\smallskip}
\hline
\end{array}
\]
\end{center}
\end{table}

\begin{figure}[t!,h]
\centering
 \includegraphics[width=0.48\textwidth]{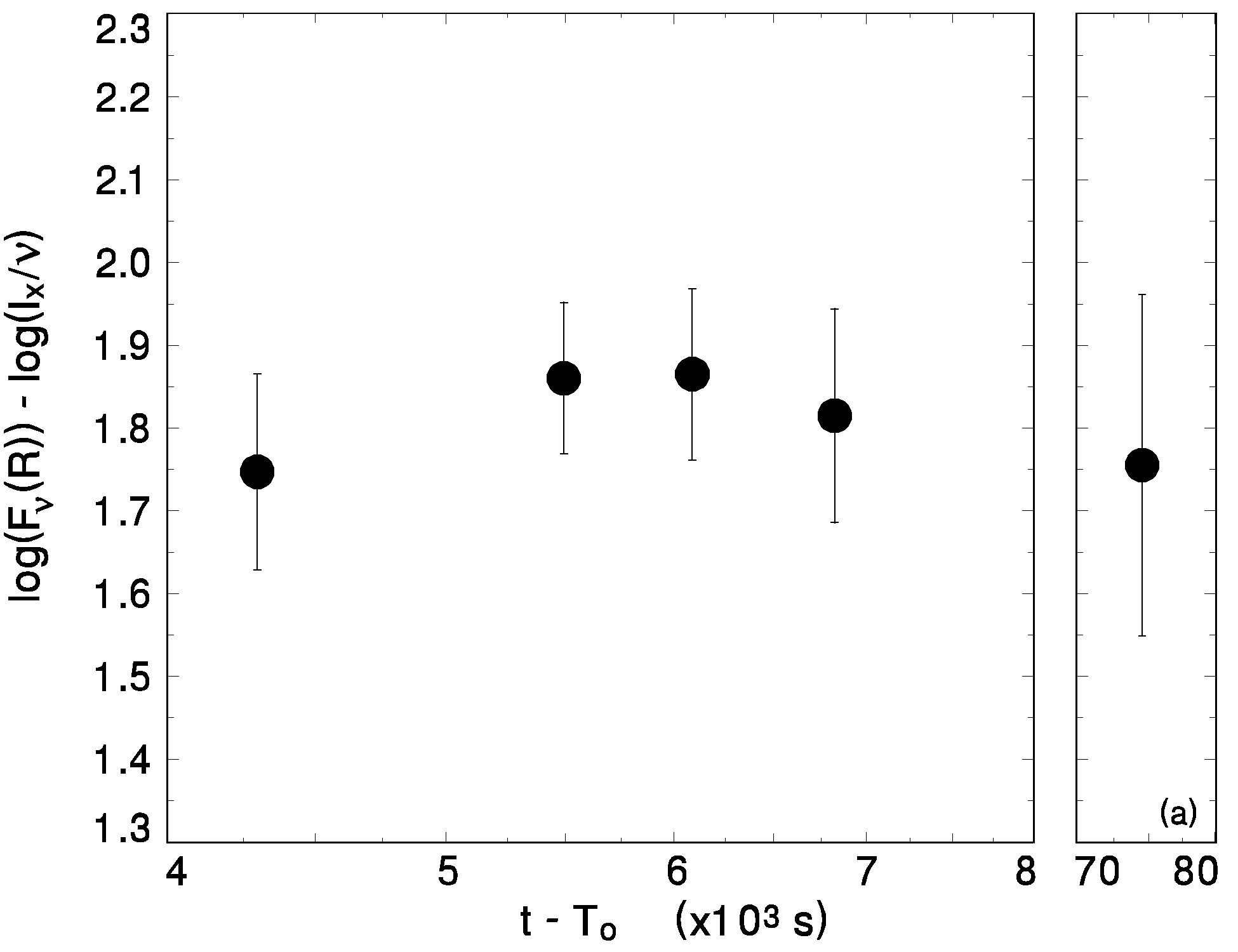} 
 \includegraphics[width=0.48\textwidth]{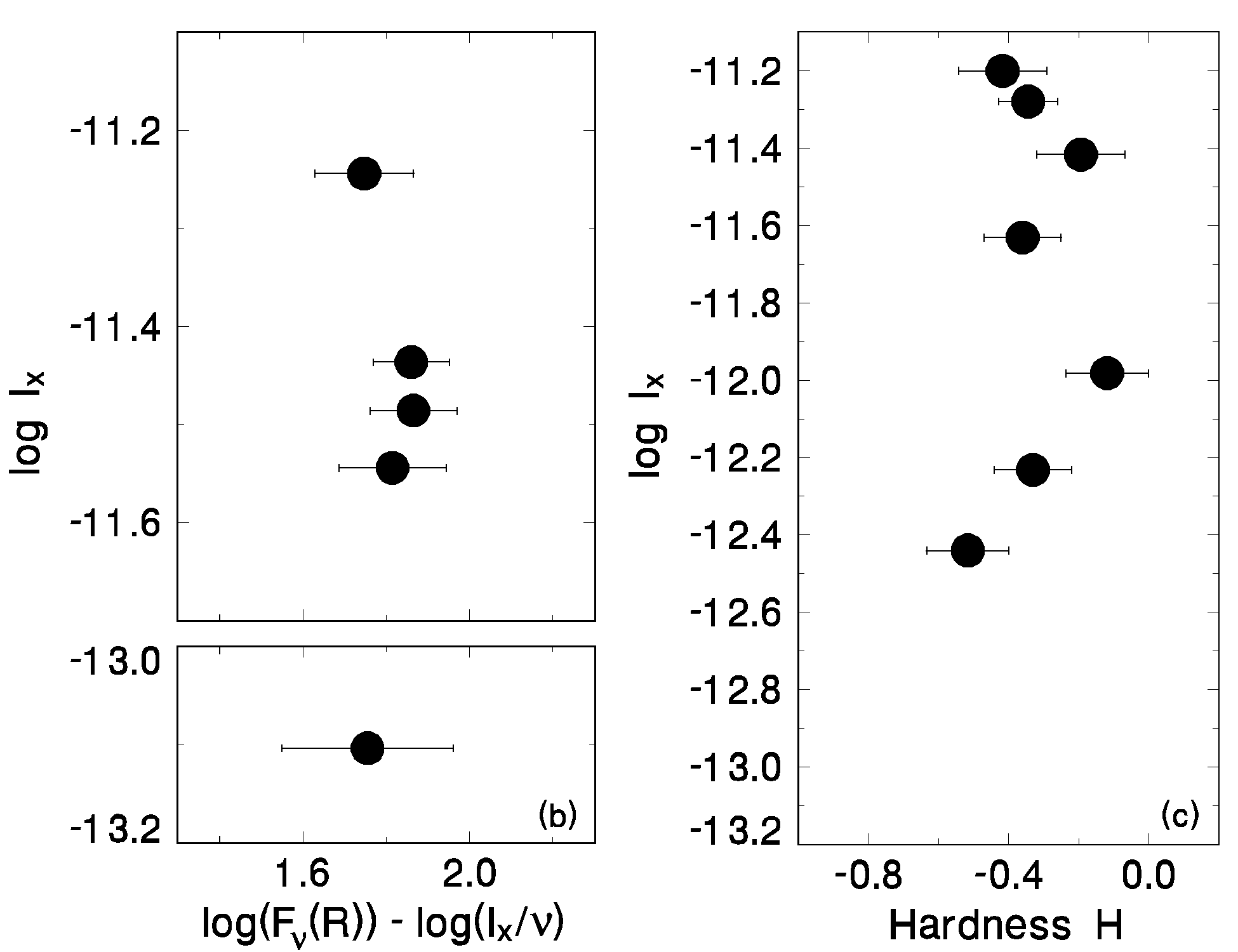} 
\caption{{\bf  (a)}  Time  evolution  of the relation of 
the  optical and  X-ray  flux   (0.3-10~keV)  in  OA  of 
GRB\,090726, defined in the text. {\bf (b)} Evolution of 
this relation with the X-ray flux. {\bf(c)} Evolution of 
X-ray  hardness  $H$  with  the  X-ray flux. The optical 
and X-ray  flux of  the  OA  were declining in all three 
diagrams. The  error  bars  were  calculated   from  the 
uncertainties given  on both  the optical and X-ray data 
at the  90~percent  level. See  Sect.\,\ref{source}  for 
details.   }
\label{opt-x}
\end{figure}

\begin{figure}
 \includegraphics[width=0.48\textwidth]{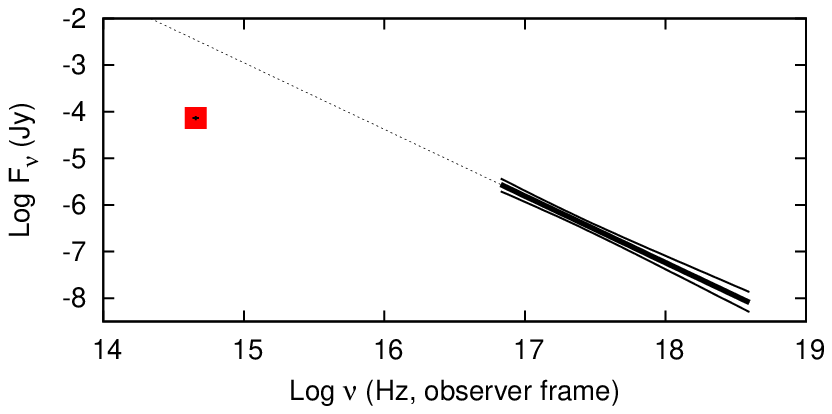} 
\caption{SED of the afterglow of GRB\,090726 at $t-T_{0}= 
4000$~s. The  X-ray data  come from  {\it Swift}/XRT, the 
optical  measurement   represents  the  extrapolation  of 
Ond\v{r}ejov data. See Sect.\,\ref{source} for details.   }
\label{sed}
\end{figure}

    An example of our CCD image of the OA, obtained at $t-T_{0}\approx670$~s, 
that is near its peak magnitude, is displayed in Fig.\,\ref{CCD}. Only a part 
of  the original  20$\times$20~arcmin  CCD  image is shown. The OA is clearly 
defined as a new, variable source. 

     The resulting  light  curve  of the  OA is displayed in Fig.\,\ref{opt}. 
The $R$ band observations of  Moskvitin  et  al. (2009)  and  Volnova et  al. 
(2009) were  also  included. Notice  that the data by Moskvitin et al. (2009) 
fit our  unfiltered  observations  nicely. The optical transient was observed 
with a 60\,cm  telescope  at Crni Vrh (Maticic \& Skvarc 2009) simultaneously 
with our measurements. We found that their light curve shows a large scatter. 
The main  contribution  of their  data is that their first point was obtained 
before the start of our observations ($t-T_{0}=154$~s vs. $t-T_{0} = 194$~s). 
The  brightness  was  fainter  than  or  comparable  to  that  of  our  first 
observation, which helps constrain any possible large-amplitude initial flash.

     In order  to  resolve  the  profile  of the early phase of OA in detail, 
both the panels  with  a linear and  log-log  scales of the axes are shown in 
Fig.\,\ref{opt}. 

     In order to show the  profile of  the early phase of OA more clearly and 
to lower the  scatter of  the  observations, Ond\v{r}ejov data were fitted by 
the code HEC13 written by Harmanec (1992). The code is based on the method of 
Vondr\'{a}k (1969, 1977), who  improved  the  original  method  of  Whittaker 
(Whittaker \& Robinson 1946). Full  description of the method can be found in 
Vondr\'{a}k (1969). This  method can fit a smooth curve to the data no matter 
what their  profile is. These  data can be distributed in time non-uniformly. 
The  method is based on minimizing the value $Q=F + \lambda^{2}S$, where $F = 
\sum p(y_{\rm i} - y'_{\rm  i})^{2}$  denotes  the  degree  of smoothing ($y$ 
being the  smoothed  and  $y'$ the observed value of the variable), $S = \sum 
(\Delta^{3}y_{\rm i})^{2}$  is  the  measure  of  roughness   of  the  curve, 
$\lambda^{2}$  is  a constant  to be  selected and defines how much the curve 
will be smoothed. HEC13  makes  use  of two  input parameters, $\epsilon$ (in 
dimensionless  units) and $\Delta T$. The quantity $\epsilon = 1/\lambda^{2}$ 
determines  how ``tight" the fit will be, that is if only the main profile or 
also the high-frequency variations are to be reproduced. The quantity $\Delta 
T$ is  the  interval  over  which the data are binned  before  smoothing. The 
resulting  fit  consists  of the  mean  points, calculated  to the individual 
observed points of the  curve. A set of fits  to the  data with the different 
input  parameters  $\epsilon$ and  $\Delta T$  was  generated  and  submitted 
to  inspection. Only  the  fit  to  the  plateau  and the decay  is shown  in 
Fig.\,\ref{opt}, because  the  transition  between  the rising branch and the 
plateau was proved to be too sharp to give any reliable fit. Also the  narrow 
flare  at  $t-T_{0} \approx 500$~s was omitted from the fitting. It was found 
that the  fit  with  $\epsilon = 1$, $\Delta T = 400$~s reproduces  the  main 
features of the light curve and significantly enhances the visibility  of the 
main profile. The residuals of this fit are displayed in Fig.\,\ref{opt}c and 
show  that  four  points, belonging to the flare and not included in the fit, 
clearly  deviate from  the  profile of this curve. This fit also allows us to 
conclude  that  not  other flare with the amplitude larger than the errors of 
the  data  occurred during  Ond\v{r}ejov  observations. It  is true that this 
method is somewhat subjective but it enables us to  find a compromise between 
a curve running through all the observed values and an ideal smooth curve. We 
preferred  to use this method because it  does not make any assumptions about 
the profile of the fitted data. 

   The relation of the optical and X-ray flux, and the evolution of the X-ray 
hardness  ratio were  investigated using {\it Swift}/XRT (Evans et al. 2007, 
2009) data  of the  afterglow  of  GRB\,090726. Only  the  $R$  band  data of 
Moskvitin et al. (2009) were used because Ond\v{r}ejov observations must have 
already finished. In order to get an analogy to the color index, the quantity 
$RX =$ log[$F_{\nu}(R)$]$-$log[$I_{\rm  X}/\nu$], where  $F_{\nu}(R)$ is  the 
flux  in  the  $R$  band, $I_{\rm  X}$ is  the  observed  X-ray  flux  in the 
0.3-10~keV band, and $\nu$ is the frequency, was  calculated. The  conversion 
between  the  $R$  band magnitude and flux was carried  out according to  the 
relation of Clark et al. (2000). The  value of  $\nu = 4\times10^{17}$~Hz was 
used for $I_{\rm X}$. Also hardness $H$ of the X-ray spectrum was  determined 
from  the  countrates  $cr$  in the hard  and  soft  bands as follows: $H = $ 
log($cr_{\rm  hard}$(1.5-10~keV))$ - $log($cr_{\rm  soft}$(0.3-1.5~keV)). The 
results  are displayed  in  Fig.\,\ref{opt-x}. Notice  that both $RX$ and $H$ 
remain  constant  within  the  observational  errors  during the decay of the 
afterglow. 

     Spectral energy  distribution (SED) of  the afterglow  of GRB\,090726 is 
displayed in  Fig.\,\ref{sed}. The  decay  slope  of  Ond\v{r}ejov  data  for 
$t-T_{0}>1730$~s was  extrapolated to $t-T_{0} = 4000$~s, that is to the time 
for which  {\it Swift}/XRT  observations exist. The X-ray data were corrected 
for the absorption given by Page  et al. (2009). The  $R$  band magnitude was 
corrected  for the  Galactic reddening by $A_{R} = 0.11$~mag according to the 
maps by Schlegel et al. (1998).

\section{Results}    \label{dis}

     We report  on a detection  of early, rising OA of GRB\,090726. Its light 
curve displayed a complicated profile, which can be summarized as follows: 

\noindent
1) Steep  rise  until  $t-T_{0}  \approx 300$~s  which  finished  by  a rapid 
transition  to a plateau lasting until $t-T_{0}\approx800$~s. 

\noindent
2) A short  flare on  the  flat  top (plateau) of  the optical light curve at 
$t-T_{0}\approx500$~s,

\noindent
3) Only a slowly steepening decline in 800~s $ < t-T_{0} < $ 1400~s. 

\noindent
4) The power-law decay steepened at $t-T_{0} \approx 1400$~s and lasted until 
at least till $t-T_{0} \approx 7000$~s. 

     Another break in the decay occurred between $t-T_{0} \approx 7000$~s and 
$t-T_{0}\approx0.86$~d (see Volnova et al. 2009).

 The initial rise for $t-T_{0}<300$~s (the first 4 points in Fig.\,\ref{opt}) 
can be approximated  by a power-law with the index $\alpha_{\rm r} = 0.94 \pm 
0.05$. This small $\alpha_{\rm r}$  speaks  in favor  of  propagation  of the 
blast wave in a wind of the progenitor rather than in the interstellar medium. 
Since the observed part of  the rise of OA occurred well after the end of the 
gamma-ray emission, this is  a thin  shell expansion (e.g. Zhang \& Kobayashi 
2005). 

     A short flare, superimposed on the flat top of the  optical light curve, 
attained the peak magnitude at $t-T_{0}\approx 500$~s. A power-law fit yields 
$\alpha_{\rm rF} = 2.7 \pm 0.8$ for the rising branch, while the decay can be 
fitted with $\alpha_{\rm dF}= -3.6\pm0.8$. This decay is considerably steeper 
than the later power-law one. Prompt emission of GRB is ruled out because the 
flare  occurred  well  after the  end of  the  observed  GRB emission. Also a 
density  enhancement  is  less  likely  because  the  decline  shallower than 
$t^{-(2 + \beta)}$ is  expected  (Fenimore  et  al. 1996). Assuming  a common 
spectral  index  $\beta = 1$  (Nardini et al. 2006), $\alpha_{\rm dF}$ should 
be $\leq-3$ in this case. If this  flare  is due  to  reverse  shock, then it 
sorts  OA of GRB\,090726 to Type~II, defined by Zhang et al. (2003). However, 
the observed decay appears to be too steep  because $\alpha_{\rm dF}\approx2$ 
is expected (Kobayashi 2000), although a wind could help steepen this decline 
(Greiner et  al. 2008). In  this regard, we  note  that any additional flash, 
like the  one in  GRB\,990123  (Akerlof et  al. 1999)  and  caused by another 
reverse shock, could occur before the start of Ond\v{r}ejov observations only 
if its peak magnitude  were  comparable with or fainter than that of the peak 
observed at $t-T_{0}\approx500$~s, The  absence of this flash is strengthened 
by the fact that also the first point of Maticic  \& Skvarc (2009) shows that 
the OA was not  brighter  than in the subsequent Ond\v{r}ejov data. The steep 
decay of the observed flare speaks in favor of activity of the central engine 
(Fan et al. 2008). This flare occurred during the phase in which OAs observed 
in X-rays  often display  a plateau  and/or flares (e.g. Burrows et al. 2005; 
Willingale et al. 2007). A flare  can  appear if the central engine continues 
to create  shells  of much  smaller power at late times with smaller $\Gamma$ 
even after GRB (Ghisellini et al. 2007).

     The profile of OA of GRB\,090726 near the peak  magnitude can be used to 
discriminate between  various  profiles of  the  jet  and the viewing angles, 
using the  recent models by Panaitescu \& Vestrand (2008). The top of this OA 
is strikingly  different  from the sharp ones of GRB\,060418 and GRB\,060607A 
(Molinari  et al. 2007; Nysewander  et al. 2009). Its flat top speaks against 
isotropic outflow  or  a jet  seen face-on. Jet with a sharp angular boundary 
with  $\Theta_{\rm obs}/\Theta_{\rm jet}>2$ can give rise to a short plateau, 
but the time of its peak luminosity $t_{\rm peak}$ occurs at $t-T_{0}>1000$~s 
for $z = 2.71$, which  is too late to be consistent  with our data. Power-law 
outflow  with $\Theta_{\rm obs}/\Theta_{\rm c}>2.5$ ($\Theta_{\rm  c}$  being 
angular size of the core) is the best case for OA of GRB\,090726. It yields a 
flat top at $t-T_{0}$  of the  order of several hundred seconds. The flat top 
thus can  be caused  by observing  the power-law jet of the OA of GRB\,090726 
off-axis. 

     Some OAs were  observed to display a re-brightening or a plateau between 
rest  frame  $(t-T_{0})_{\rm  rest}$  of  0.01  and  0.02~d  (i.e.  $t-T_{0}$ 
between $\sim$3200  and  $\sim$6400~s for $z=2.71$) (Klotz et al. 2005). This 
time  period  corresponds  to  the  phase  of  the  power-law  decay in OA of 
GRB\,090726. Also  the  X-ray 0.3-10~keV data display only  smooth, power-law 
decay without any plateau in this epoch. 

     Determination of the initial Lorentz factor $\Gamma_{0}$ of GRB requires 
a knowledge of its isotropic  energy  release  $E_{\rm  iso}$.  Since  only a 
15-150~keV  spectrum, best  fitted by  a simple  power-law model (Page et al. 
2009), is available, we estimated $E_{\rm iso}$ from Amati relation (Amati et 
al. 2002). The observed spectrum suggests $E_{\rm iso} > 2\times 10^{52}$~erg. 
The upper limit of GRBs is  $E_{\rm iso}\approx3\times10^{54}$~erg. Following 
the  approach  of  Molinari  et  al. (2007), the  initial  Lorentz  factor of 
GRB\,090726 turns out to be $\Gamma_{0}\approx230-530$ in case of propagation 
of the  blast  wave  in  a homogeneous medium, while propagation of this wave 
in  a wind  environment gives  $\Gamma_{0} \approx 80-300$. The  density  $n= 
1$~cm$^{-3}$ and the radiative  efficiency  $\eta = 0.2$ were assumed in both 
cases. Taking  $t_{\rm  peak} = 300$~s  or 500~s  has  only  minor  effect on 
$\Gamma_{0}$ in our case because the uncertainties in $E_{\rm iso}$ dominate. 
Rykoff  et al. (2009) found  $\Gamma_{0}$ between $\sim$100 and $\sim$1000 in 
their ensemble of GRBs. The value  of $\Gamma_{0}$ in  GRB\,090726 thus falls 
into the lower half of this range and, given  the uncertainty  in its $E_{\rm 
iso}$, it may even lie on the lower end.

     Ond\v{r}ejov data show that the optical light curve achieves a power-law 
decay with $\alpha_{\rm dO}=-1.02\pm0.12$ from $t-T_{0}= 1730$~s till the end 
of the  series, which  is a typical  value for OAs. The X-ray observations of 
the  afterglow  of  GRB\,090726  started  when  the  OA  emission was already 
declining  (see  Fig.\,\ref{opt}bd). The decline  of  the  X-ray emission was 
power-law with  $\alpha_{\rm dX} = -1.27\pm0.09$ before a break at $t-T_{0} = 
51^{+18}_{-7}$~ks (Page et al. 2009). Both $\alpha_{\rm dO}$ and $\alpha_{\rm 
dX}$ can be regarded as identical  at  $\sim2\sigma$  level. The ratio of the 
optical  and  X-ray  flux can be considered constant (Fig.\,\ref{opt-x}). All 
this suggests that both the optical and X-ray emission decayed simultaneously 
and  that the spectral  profile  from X-ray to the optical band did not vary. 
This is true for  both the  time  period  before  and  after the break in the 
decaying  light curve; no significant flares were observed in the data set of 
the 0.3-10~keV band. The SED in Fig.\,\ref{sed} suggests  a break between the 
optical  and  X-ray band, with the X-ray part having a steeper slope than its 
extrapolation to  the optical  band, in the  same sense as observed for OA of 
GRB\,060418 by Molinari et al. (2007). Correction of the $R$ band  point by a 
very  high  $A_{R}  \approx  3.7$~mag  due  to  the  extinction  in the  host 
galaxy  would  make it  lie on  the  extrapolation  of  the X-ray spectrum in 
Fig.\,\ref{sed}. However, the  relation  between  the optical  extinction and 
X-ray absorption in the host is unknown. 

     The break in  the  X-ray  light curve occurred in the gap of the optical 
data, but the  ratio  of the  optical  and  X-ray  flux  before and after the 
break shows that this event is consistent with an achromatic break. Following 
the approach  of  Ghirlanda et  al. (2004), the Lorentz factor at the time of 
this break has already decreased to $\Gamma_{\rm break}\approx20$. 

     Although the  intrinsic  absorbing  column  of  the X-ray  afterglow  of 
GRB\,090726 is larger  than the  Galactic  one (Page 2009), no changes of its 
$H$ were observed  during  its power-law decline covering the part before the 
break in  its  light  curve  (Fig.\,\ref{opt-x}).  This  suggests  a constant 
absorption  which  does  not participate  in the evolution of this afterglow. 
This can be explained  if this absorption comes from the regions more distant 
from this GRB, e.g. if this event is behind  a molecular  cloud. Any possible 
variations of the absorption caused by emission of this GRB  or its afterglow 
must  have  occurred before the start of XRT observing, that is at $t-T_{0} < 
3000$~s. The  extinction of the  optical emission depends on the dust content 
while the  absorption  of  X-rays  depends  on  the  gaseous component of the 
absorbing medium. If  only the dust is being destroyed during the early phase 
of the afterglow or condensed during  its  later phase, then $RX$ would vary, 
which is not  observed in our case (Fig.\,\ref{opt-x}). The constancy of both 
$RX$  and  $H$  (within the observing scatter) suggests that neither the dust 
nor the gaseous component underwent any evolution during this OA event. 

     We find that the power-law decay rate of OA of GRB\,090726 is comparable 
to that of the ensemble of OAs with homogeneous optical  spectra at $z=0.36 - 
3.5$ (\v{S}imon  et  al. 2001, 2004). The  $k$-corrected  absolute  magnitude 
(\v{S}imon et al. 2001) $M_{\rm  R_{0}} \approx -22$  of OA of GRB\,090726 at 
$(t-T_{0})_{\rm rest}\approx0.2$~d places it below or at the lower end of the 
distribution of this ensemble of  OAs. $M_{\rm  R_{0}}$ of  the ensemble span 
about  4~mag  at  this  $(t-T_{0})_{\rm  rest}$. This  is consistent with the 
synchrotron  emission of the GRB afterglow fireball emission model by Sari et 
al. (1998) with very similar spectra but very different luminosity at a given 
$(t-T_{0})_{\rm  rest}$. In this phase, OA of GRB\,090726 thus belongs to the 
least luminous ones and is similar to OA of X-ray flash XRF\,030723 (Fynbo et 
al. 2004; \v{S}imon et al. 2006).

\begin{acknowledgements}
The  support by the grants 205/08/1207 and 102/09/0997 of the Grant Agency of 
the Czech Republic and the project PECS 98023 is acknowledged. This work made 
use of data supplied by the UK Swift Science Data Centre at the University of 
Leicester. We thank Prof.~Harmanec  for  providing us with the program HEC13. 
The  Fortran  source  version, compiled version and brief instructions how to 
use the program can be obtained via ftp://astro.troja.mff.cuni.cz/hec/HEC13. 
\end{acknowledgements}

\end{document}